\documentclass[conference]{IEEEtran}
\makeatletter
\def\ps@headings{
\def\@oddhead{\mbox{}\scriptsize\rightmark \hfil \thepage}
\def\@evenhead{\scriptsize\thepage \hfil \leftmark\mbox{}}
\def\@oddfoot{}
\def\@evenfoot{}}
\makeatother
            
\usepackage[letterpaper,
            left=0.625in,
            right=0.625in,
            top=0.75in,
            bottom=1in]
            {geometry}

\usepackage{epsfig,latexsym,amsmath,epsf,amssymb,amsfonts,bm,theorem,subfigure,epstopdf,cite,authblk, bbm, color}

\DeclareMathOperator*{\argmaxA}{arg\,max}

\begin{document}

\title{Free-Rider Games for Federated Learning with Selfish Clients in NextG Wireless Networks}
	
\author[]{Yalin E. Sagduyu}
\vspace{-0.5cm}
\affil[]{\normalsize  National Security Institute, Virginia Tech, Arlington, VA, USA \\ Email: ysagduyu@vt.edu}
\maketitle
\begin{abstract}
This paper presents a game theoretic framework for participation and free-riding in federated learning (FL), and determines the Nash equilibrium strategies when FL is executed over wireless links. To support spectrum sensing for NextG communications, FL is used by clients, namely spectrum sensors with limited training datasets and computation resources, to train a wireless signal classifier while preserving privacy. In FL, a client may be free-riding, i.e., it does not participate in FL model updates, if the computation and transmission cost for FL participation is high, and receives the global model (learned by other clients) without incurring a cost. However, the free-riding behavior may potentially decrease the global accuracy due to lack of contribution to global model learning. This tradeoff leads to a non-cooperative game where each client aims to individually maximize its utility as the difference between the global model accuracy and the cost of FL participation. The Nash equilibrium strategies are derived for free-riding probabilities such that no client can unilaterally increase its utility given the strategies of its opponents remain the same. The free-riding probability increases with the FL participation cost and the number of clients, and a significant optimality gap exists in Nash equilibrium with respect to the joint optimization for all clients. The optimality gap increases with the number of clients and the maximum gap is evaluated as a function of the cost. These results quantify the impact of free-riding on the resilience of FL in NextG networks and indicate operational modes for FL participation.
\end{abstract}
\section{Introduction}\label{sec:Introduction}
\emph{Federated learning} (FL) allows a set of clients to collectively train a global machine learning model without sharing their individual training datasets \cite{McMahan17:FL}. In orchestration by a server, each client in FL shares its trained local model with the server and the server aggregates these models and shares the aggregated global model back with the clients. Then, the clients initialize their local models with this global model and train these models using their own training data samples. By repeating this process over multiple epochs, a global model is trained in this client-server framework. 
 FL provides various benefits: (i) Since the training datasets are not shared in FL, clients can preserve the privacy of their training datasets. (ii) Since the trained models are typically smaller than the training data, the communication load is reduced in FL compared to the case of sharing the data among clients and the server. (iii) Clients that may not have large training datasets or powerful computing resources can still collectively train a global model in FL by using their local training datasets and computing resources 
\cite{Survey1:FL, Survey3:FL, Survey0:FL}.

\begin{figure}[ht]
\vspace{-0.4cm}
  \centering
  \includegraphics[width=0.65\columnwidth]{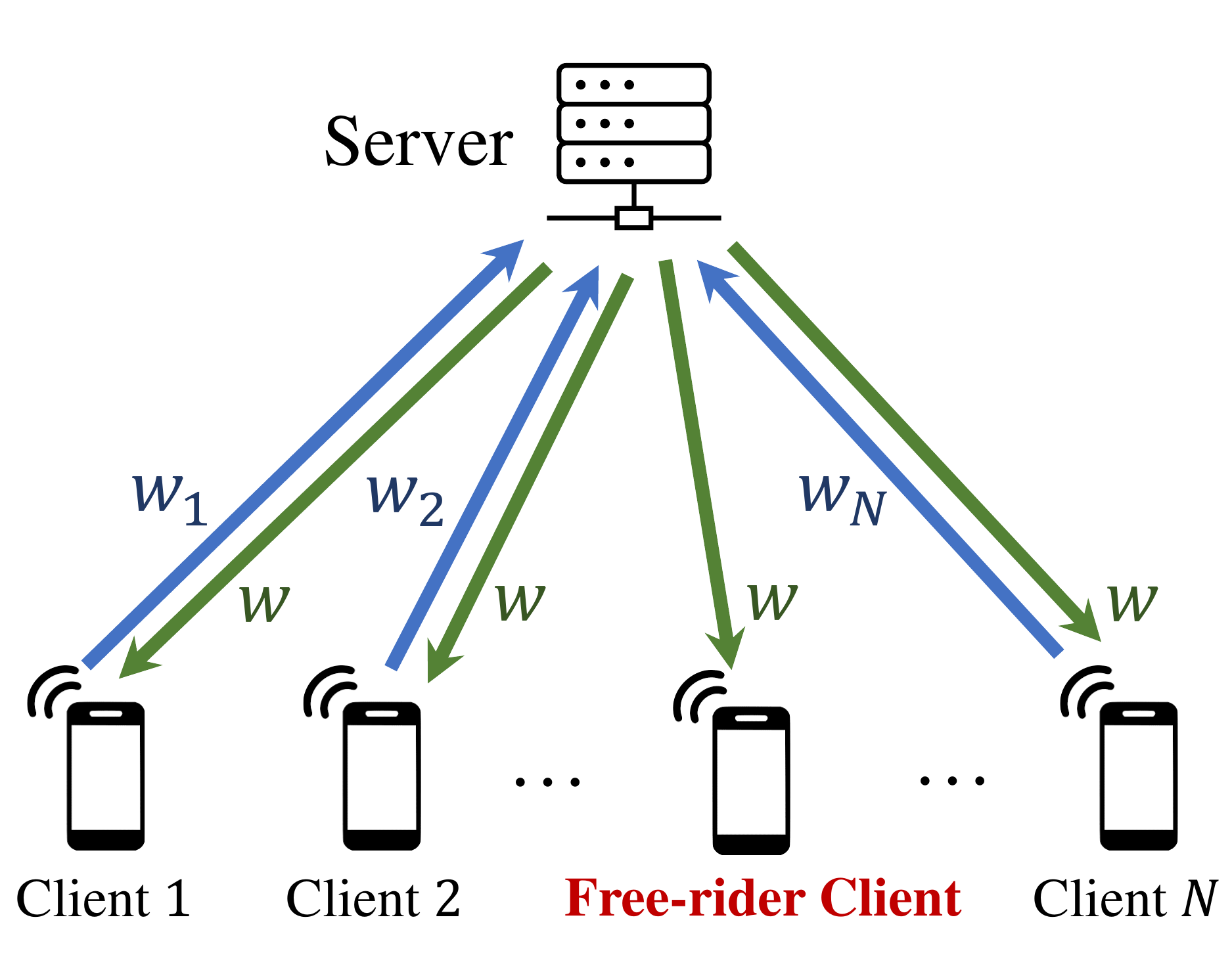}
  \vspace{-0.2cm}
  \caption{Clients are either participating in FL model updates or free-riding ($w_i$ and $w$ correspond to the weights of client $i$'s local model and global model).}
  \label{fig:FLfreeriding}
\end{figure}

To reap these benefits of FL, clients may need incentives to participate in FL. Various incentive mechanisms such as those motivated by game theory have been studied to foster collaboration for FL \cite{Game1:FL, Game2:FL}. Also, it is possible for clients to decide with whom they participate in FL by forming federating coalitions to balance the performance of local and global models \cite{Coalition:FL}. On the other hand, FL
is also susceptible to various insider exploits \cite{FLsecurity1, FLsecurity2, FLsecurity3, FL:MEC} such as (i) data poisoning
(training datasets are manipulated by malicious clients), (ii) model update poisoning (models are manipulated by malicious clients before sharing with the server),  and (iii) inference of class representatives, memberships, and training inputs and labels. One particular client behavior of concern is \emph{free-riding} \cite{Freerider1, Freerider2}, where some clients may not contribute to the FL model updates, but they may receive the global model from the server (as shown in Fig.~\ref{fig:FLfreeriding}). While there may be a benign reason that the free-riding client does not have training data to contribute to the FL model updates, there may be also selfish reason that this client may avoid incurring the cost of spending its communication and computation resources, or a malicious reason that this client may aim to steal the global model.  

Wireless systems can support various applications of FL such as in mobile edge networks \cite{FL:MEC}, Internet of Things (IoT) \cite{FL:IoT}, 5G \cite{FL:5G, FL:5G-2}, and 6G \cite{FL:6G}. FL can be performed over wireless links \cite{Wireless1:FL, Wireless2:FL, Wireless3:FL, Wireless5:FL, Wireless6:FL} and multi-hop wireless networks \cite{Network2:FL}. In these cases, clients may refrain from participating in the FL model updates to preserve their energy that would be spent for wireless transmissions to share the FL model updates (especially when clients are battery-operated).  The cost for participating in FL increases linearly with the number of epochs used in FL, where each client participating in FL spends energy and resources for locally training its deep neural network model as well as for transmitting its model update back to the server as an uplink transmission in each epoch. On the other hand, a free-riding client does not incur these costs, as it only receives the model updates from the server without any need of training the local model and transmitting the model updates to the server.
\vspace{0.02in}

While free-riding clients avoid these costs, their selfish behavior adversely affects the global server accuracy and consequently the performance of other clients. In particular, the gains from getting information from the server (namely, the global server accuracy) is reduced when some clients are not participating in FL, as they do not contribute to the global model learning with their individual data. Thus, there is a trade-off that leads to the conflicting interests of selfish clients when they assume the roles as participating in FL model updates (called `participant') or free-riding without participating in FL model updates (called `free-rider'). 
\vspace{0.02in}

This paper formulates these conflicting interests as a \emph{non-cooperative free-rider game} among clients involved in FL. Each client is a player and individually selects to be a participant or a free-rider as its action. The strategy of each client is the probability of selecting to be a free-rider. The reward is the global server accuracy and a cost is associated with participating in FL model updates. The utility (payoff) of each client is computed as the difference of the reward and the cost, and it is individually maximized given the strategies (free-riding probabilities) of other clients.   
\vspace{0.02in}

In this paper, FL is used to collectively train a \emph{wireless signal classifier}. A real-world scenario is the 3.5GHz Citizens Broadband Radio Service (CBRS) \cite{FCC},  where the incumbent user (namely, radar) and 5G communications co-exist in the same spectrum band. The spectrum sensors of the Environment Sensing Capability (ESC) need to detect the incumbent signal and inform the Spectrum Access System (SAS) to configure 5G communications and prevent interference with the incumbent signal. Spectrum sensors can use a deep learning model to detect signals of interest \cite{DL:CBRS, DL:CBRS-2}. Each sensor may have limited training data and processing power, and may not share the data to keep communication load low and/or preserve data privacy (as spectrum sensors may monitor different geographical areas with different levels of privacy requirements). Thus, FL can be applied to train the deep learning model by utilizing training datasets collected at different locations  \cite{Wireless6:FL, Network2:FL}. 
\vspace{0.02in}

In this wireless network setting, the \emph{pure and mixed strategies} are derived for free-riding in \emph{Nash equilibrium}, where no client can unilaterally improve its utility by changing its strategies given the strategies of other clients are fixed. The free-riding probabilities in Nash equilibrium increase with the cost and the number of clients involved in FL participation. In the meantime, the corresponding utility in Nash equilibrium is measured and the optimality gap is evaluated relative to the case when the total utility is maximized by selecting the free-riding probabilities jointly. The optimality gap increases with the number of clients and the cost for which the optimality gap reaches its maximum is identified. Also, the \emph{fictitious play} is considered, where each client plays the best response to the empirical distributions of opponent strategies over time (without knowing their utilities).
\\
\indent Overall, the results presented in this paper quantify the resilience of FL to free-riding in a non-cooperative setting, when FL is applied to the wireless signal classification task, and indicate operational modes for clients to participate in FL model updates, when FL is performed over wireless links with associated costs for computing and transmitting. 
\\
\indent 
The rest of the paper is organized as follows. Section~\ref{sec:SystemModel} describes the system model. Section~\ref{sec:Nash} presents the Nash equilibrium strategies for the free-rider game. Section~\ref{sec:Wireless} studies the free-rider game when FL is applied for wireless signal classification. Section~\ref{sec:Conclusion} concludes the paper. 

\section{System Model}\label{sec:SystemModel}
There are $N$ clients from the set $\mathcal{N}$ that may participate in FL model updates. Each participating client $i \in \mathcal{N}$ has its own training data, trains its own local model (weights $w_i$) in each epoch of FL, and sends it to the server. The server builds the global model by the federated averaging (FedAvg) algorithm such that the weights of the global model are aggregated  as $w = \frac{1}{N} \sum_{i \in \mathcal{N}} w_i$. Then, the server broadcasts this global model $w$ to all clients in one transmission (again over the air). Clients update their local models by this global model $w$ (i.e., they initialize their models with $w$). This set of operations corresponds to one epoch of FL and is repeated over many epochs to build a global model. The over-the-air transmissions between the clients and the server are separated over time or frequency so that they do not interfere with each other.

It is possible that one or more clients from $\mathcal{N}$ are \emph{free-riding}, i.e., they do not participate in FL model updates by refraining to send their model updates to the server, as shown in Fig.~\ref{fig:FLfreeriding}. However, they still receive the global model update from the server (that is broadcast in the wireless medium). With participation in FL model updates, a client may incur a cost for consuming energy for computation and transmission, while it may end with higher global model accuracy by contributing to the model updates. With free-riding, a client does not incur a cost, but may end up with lower accuracy as it does not contribute to the aggregate model of the server. To formulate this tradeoff, a \emph{game} is set up among \emph{selfish clients} where their actions are to select between two roles, namely free-riding and participating in FL model updates.

The strategy of each client $i$ is to select its free-riding probability $p_i$. Let ${\bf{p}}_{-i}$ denote the strategies (namely, free-riding probabilities) of $\mathcal{N}_{-i}$, where $\mathcal{N}_{-i} = \mathcal{N} - \{i\} $ is the set of clients excluding client $i$. Each client $i$ receives the global server model accuracy as the reward. Specifically, client $i$ receives the reward  $r^{(i)}_{S_i|{\bf{S}}_{-i}}$, if client $i$ selects the role $S_i \in \{F, P\}$, where $F$ and $P$ correspond to `free-rider' and 'participant' roles, respectively, and other clients select roles ${\bf{S}}_{-i} = \left\{ S_j\right\}_{j \in \mathcal{N}_{-i}}$. Suppose that $\mathcal{N}_{-i}$ is partitioned into sets $\mathcal{F}$ and $\mathcal{P}$ based on the roles $F$ and $P$ assumed by clients in $\mathcal{N}_{-i}$ such that $\mathcal{F} \cup \mathcal{P} = \mathcal{N}_{-i}$ and $\mathcal{F} \cap \mathcal{P} = \varnothing $. Then, the reward of client $i$ is $u^{(i)}_{F|\mathcal{F},\mathcal{P}}$ if $S_i = F$, and $u^{(i)}_{P|\mathcal{F},\mathcal{P}}$ if $S_i = P$. In the latter case, client $i$ also incurs a computation and transmission cost $c_i$. 

\section{Nash Equilibrium Strategies for the Free-Rider Game} \label{sec:Nash}
\noindent {\bf Utility functions:} The utility function combines rewards and costs. In response to the strategies, ${\bf{p}}_{-i}$, of its opponents, the utility of client $i$, if it selects to be free-riding, is given by 
\begin{equation} \label{eq:uF}
    u_{F}^{(i)} \left( {\bf{p}}_{-i} \right) = \sum_{\substack{ \{\mathcal{F}, \mathcal{P}\}: \\ \mathcal{F} \cup \mathcal{P} = \mathcal{N}_{-i}, \\ \mathcal{F} \cap \mathcal{P} = \varnothing }} \prod_{j \in \mathcal{F}, k \in \mathcal{P}} p_j \left (1-p_k \right) u^{(i)}_{F|\mathcal{F}, \mathcal{P}} 
\end{equation}
and the utility of client $i$, if it selects to participate in FL model updates, is given by
\begin{equation} \label{eq:uP}
    u_{P}^{(i)} \left( {\bf{p}}_{-i} \right) = \hspace{-0.2cm} \sum_{\substack{ \{\mathcal{F}, \mathcal{P}\}: \\ \mathcal{F} \cup \mathcal{P} = \mathcal{N}_{-i}, \\ \mathcal{F} \cap \mathcal{P} = \varnothing }}  \prod_{j \in \mathcal{F}, k \in \mathcal{P}} p_j \left (1-p_k \right) u^{(i)}_{P|\mathcal{F}, \mathcal{P}}  \:\:\:\:  - c_i. 
\end{equation}

Averaged over its strategy $p_i$, the utility of client $i$ as a function of ${\bf p} = \left\{ p_i, {\bf p_{-i}} \right\}$ is given by
\begin{equation} \label{eq:uT}
    u_T^{(i)} ({\bf p}) = \sum_{i \in \mathcal{N}} \left( p_i \: u_{F}^{(i)} \left( {\bf{p}}_{-i} \right) +  (1-p_i) \: u_{P}^{(i)} \left( {\bf{p}}_{-i} \right) \right), 
\end{equation}
where $u_{F}^{(i)}\left( {\bf{p}}_{-i} \right)$ and $u_{P}^{(i)}\left( {\bf{p}}_{-i} \right)$ are given in (\ref{eq:uF}) and (\ref{eq:uP}), respectively. 

\noindent {\bf Globally optimal (cooperative) strategies:} The globally optimal strategies ${\bf p}^*$ are the ones that maximize the total utility $u_T ({\bf p})$ for all clients, namely ${\bf p}^*$ is given by
\begin{equation}
    {\bf p}^* = \argmaxA_{{\bf p}: \: p_i \in [0,1]} \:\: \sum_{i \in \mathcal{N}}  u_T^{(i)} ({\bf p}).
\end{equation}

\noindent {\bf Nash equilibrium strategies:} To compute the \emph{non-cooperative Nash equilibrium strategies}, $B_i\left({\bf{p}}_{-i} \right)$ is defined as the best response of client $i$ to the strategies ${\bf{p}}_{-i}$ of other clients $\mathcal{N}_{-i}$. Then, the Nash equilibrium strategies satisfy 
\begin{equation}
{\bf p}^* \in  B_i\left({\bf p}^*_{-i}\right), \:\: i \in \mathcal{N},
\end{equation}
where $B_i\left({\bf p}^*_{-i}\right)$ is obtained by solving the optimization problem given by
\begin{eqnarray}
&& \max_{p_i} \: \: u_T^{(i)} \left(p_i, {\bf p}_{-i}\right) \nonumber \\
&& \text{subject to } 0 \leq p_i \leq 1.\label{eq:indvopt}
\end{eqnarray}

In Nash equilibrium, no client can unilaterally deviate from its strategy to increase its utility. To find the Nash equilibrium strategies, (\ref{eq:indvopt}) is converted first to a standard optimization form:
\begin{eqnarray}
&& \min_{x} f(x) \nonumber \\ 
&& \text{subject to } g_j(x) \leq 0, \label{eq:std}
\end{eqnarray}
where $f$ is the objective function, $x$ is the optimization variable, and $g_j$ is the $j$th inequality constraint function. For client $i$, (\ref{eq:indvopt}) is converted to standard form (\ref{eq:std}) by setting 
\begin{eqnarray}
&& x = p_i, \label{eq:condfirst}\\
&& f(x) = - u_T^{(i)} \left(x, {\bf p}_{-i}\right), \\
&& g_1(x) = -x,  \\ &&   g_2(x) = x - 1 \label{eq:condlast}
\end{eqnarray}
for ${\bf p}_{-i}$, where $g_1$ and $g_2$ correspond to the constraints $p_i \geq 0$ and  $p_i \leq 1$, respectively, in the optimization problem (\ref{eq:indvopt}). 

In general, the Karush–Kuhn–Tucker (KKT) conditions for (\ref{eq:std}) are given by 
\begin{eqnarray}
&& \nabla_x f(x^*) + \sum_{j} \mu_{j} \nabla_x g_j(x^*) = 0, \label{eq:KKT2}\\
&& g_j (x^*) \leq 0, \\
&& \mu_j \geq 0, \\
&& \sum_j \mu_j g_j(x^*) = 0 \label{eq:KKT3}.
\end{eqnarray}

By applying (\ref{eq:condfirst})-(\ref{eq:condlast}) for each client $i$, the KKT conditions for the utility optimization for client $i$ are given by
\begin{eqnarray}
&& - \nabla_{p_i} u_T^{(i)} \left(p_i^*, {\bf p}_{-i}\right) - \mu_{i,1} + \mu_{i,2} = 0, \label{eq:KKT1} \\
&& p_i^* \geq 0, \:\:\: p_i^* \leq 1, \\
&& \mu_{i,j} \geq 0, \\
&& \mu_{i,1} \: p_i^* = 0, \:\:\: \mu_{i,2} \: (p_i^*-1) = 0,
\end{eqnarray}
where $\mu_{i,j}$ is the KKT multiplier for the $j$th constraint of client $i$. Note that the condition (\ref{eq:KKT1}) couples the strategies of different clients and can be expressed (by applying the utilities from (\ref{eq:uF})-(\ref{eq:uT})) as 
\begin{eqnarray} \label{eq:KKTspec}
&& \sum_{\substack{ \{\mathcal{F}, \mathcal{P}\}: \\ \mathcal{F} \cup \mathcal{P} = \mathcal{N}_{-i}, \\ \mathcal{F} \cap \mathcal{P} = \varnothing }} \prod_{j \in \mathcal{F}, k \in \mathcal{P}} p_j \left (1-p_k \right) \left( u^{(i)}_{P|\mathcal{F}, \mathcal{P}} - u^{(i)}_{F|\mathcal{F}, \mathcal{P}}  \right) \nonumber \\ && \hspace{0.5cm} = c_i + \mu_{i,1} - \mu_{i,2}
\end{eqnarray}
for each client $i \in \mathcal{N}$. Note that both pure strategies (when $p_i \in \{0, 1\}$ for all $i \in \mathcal{N}$) and mixed strategies may exist as the Nash equilibrium strategies. 

\noindent {\bf Fictitious play:} In fictitious play,  each player assumes that the opponents play stationary (possibly mixed) strategies that are given by the empirical frequency of their past actions. At each round of fictitious play, each player plays the best response to the empirical frequency of strategies chosen by their opponents until this round.  Define $p_{i}(t)$ as the strategy assumed by client $i$ at round $t$ and $\tilde{p}_{i}(t)$ as the moving average of the strategies played by client $i$ until round $t$. Then, the strategy of client $i$ is given by
\begin{equation}
p_{i}(t) = B_i(\tilde{{\bf p}}_{-i}(t))
\end{equation}
at round $t$, where $\tilde{{\bf p}}_{-i}(t) = \{\tilde{p}_{i}(t)\}_{i \in \mathcal{N}_{-i}}$. 

Note that there is no need for any client to know the utilities of other clients. At each round, the server may broadcast the list of clients participating in FL so that each client $i$ can compute $\tilde{{\bf p}}_{-i}(t)$ over time by tracking the actions of other clients. 

The strict Nash equilibrium strategies in fictitious play are absorbing states, i.e., if all players select their Nash equilibrium strategies at a given round, then they continue to select these strategies for all subsequent rounds. Note that if fictitious play converges to a distribution, this corresponds to a Nash equilibrium. 

\noindent {\bf Nash equilibrium strategies for two clients:}
The expected utility of client $i = 1,2$ for free-riding given the other client's strategy, $p_{-i}$, follows from (\ref{eq:uF}) as
\begin{equation} \label{eq:UF2}
u^{(i)}_F(p_{-i}) = p_{-i} u^{(i)}_{F|F} + (1-p_{-i}) u^{(i)}_{F|P}
\end{equation} and the expected utility of client $i$ for participating in FL given $p_{-i}$ follows from (\ref{eq:uP}) as 
\begin{equation} \label{eq:UP2}
u^{(i)}_P(p_{-i}) = p_{-i} u^{(i)}_{P|F} + (1-p_{-i}) u^{(i)}_{P|P} - c_i.
\end{equation}

The Nash equilibrium strategies for two clients are obtained by solving the KKT conditions (\ref{eq:KKT2})-(\ref{eq:KKT3}) with utilities (\ref{eq:UF2}) and (\ref{eq:UP2}). The resulting pure equilibrium strategies are given by 

\begin{equation} \label{eq:pure}
  \left( p_1^*, p_2^* \right) =
    \begin{cases}
      (0,0) & \text{if $u_{F|P}^{(i)} \leq u_{P|P}^{(i)} - c_i$, $i=1,2$},\\
      (1,1) & \text{if $u_{F|F}^{(i)} \geq u_{P|F}^{(i)} - c_i$, $i=1,2$}, \\
      (0,1) & \text{if $u_{F|F}^{(1)} \leq u_{P|F}^{(1)} - c_1$ and} \\ 
       & \hspace{0.35cm} \text{$u_{F|P}^{(2)} \geq u_{P|P}^{(2)} - c_2$,} \\
       (1,0) & \text{if $u_{F|F}^{(2)} \leq u_{P|F}^{(2)} - c_2$ and} \\ 
       & \hspace{0.35cm} \text{$u_{F|P}^{(1)} \geq u_{P|P}^{(1)} - c_1$}.
    \end{cases}
\end{equation}

Depending on the rewards and costs of clients, it is possible that one client selects a deterministic strategy and the other client selects a randomized strategy, or both select randomized strategies. These mixed strategies in Nash equilibrium are computed as 
\begin{equation} \label{eq:mixed}
  \left( p_1^*, p_2^* \right) =
    \begin{cases}
      (0, p) & \text{if $u_{F|P}^{(2)} = u_{P|P}^{(2)} - c_2$},\\  & p \geq p_{1,M}, p \in [0,1], \\
        (p, 0) & \text{if $u_{F|P}^{(1)} = u_{P|P}^{(1)} - c_1$},\\  & p \geq p_{2,M}, p \in [0,1],  \\
        (1, p) & \text{if $u_{F|F}^{(2)} = u_{P|F}^{(2)} - c_2$},\\ & p \leq p_{1,M}, p \in [0,1],  \\
        (p, 1) & \text{if $u_{F|F}^{(1)} = u_{P|F}^{(1)} - c_1$},\\ & p \leq p_{2,M}, p \in [0,1], \\
        (p_{1,M}, p_{2,M}) & \text{otherwise},
    \end{cases}
\end{equation}
where $p_{i,M}$, $i=1,2$, follows by solving (\ref{eq:KKTspec}) with $\mu_{i,j} =0$, $i=1,2$ and $j=1,2$, and is given by  
\begin{eqnarray} \label{eq:pim}
p_{i,M} =  \left[ \frac{c_{-i} + u^{(-i)}_{F|P} - u^{(-i)}_{P|P} }{ u^{(-i)}_{P|F} - u^{(-i)}_{P|P} + u^{(-i)}_{F|P} - u^{(-i)}_{F|F}} \right]^1_0, 
\end{eqnarray}
for $i=1,2$, where 
\begin{equation}
  \left[ x \right]^1_0 =
    \begin{cases}
      0 & \text{if $x < 0$},\\
      x & \text{if $0 \leq x \leq 1$},\\
      1 & \text{otherwise.}
    \end{cases}
\end{equation}
\noindent {\bf Illustrative example:} Consider an example of two clients. FL is used by these two clients to train a binary classifier. The reward of each client is the accuracy of the learned global model.  Without any clients participating in FL, $u^{(i)}_{F|F} = 0.5$ for $i=1,2$, namely the default accuracy for the global model is $0.5$ (corresponding to the case of flipping a fair coin). Consider the symmetric case that follows by assuming symmetric costs $c_i =c$ and symmetric rewards
$u^{(i)}_{P|P} = u_{P|P} = 0.9$, and $u^{(i)}_{P|F} =  u^{(i)}_{F|P} = u_{F|P} = 0.8$,  $i=1,2$. From (\ref{eq:pim}), $p_{i,M} = p_{M} =  \left[5(c-0.1)\right]_0^1$, $i=1,2$. Then, the Nash equilibrium strategies are computed from (\ref{eq:pure})-(\ref{eq:mixed}) as 
\begin{equation}
  \left( p_1^*, p_2^* \right) =
    \begin{cases}
      (0,0) & \text{if $ c \leq 0.1$}\\
      (0,p), (p,0) & \text{if $ c = 0.1$}, \\ & p \geq p_{1,M}, p \in [0,1],\\
      5(c-0.1)& \text{if $0.1 \leq c \leq 0.3$}\\
      (1,0), (0,1) & \text{if $0.1 \leq c \leq 0.3$}\\
       (1,p), (p,1) & \text{if $ c = 0.3$}, \\ & p \leq p_{1,M}, p \in [0,1], \\
      (1,1) & \text{if $c  \geq 0.3$}.
    \end{cases}
\end{equation}

Note that as $c$ increases, the free-riding probability of clients increases and the total utility drops until it saturates (when the free-riding probability becomes $1$). The optimality loss is the highest when $c = 0.3$ such that the free-riding probability is $1$ in Nash equilibrium and $0.25$ in the globally optimal case. In this case, the optimality loss of total utility in Nash equilibrium (relative to the globally optimal case) is computed as $\frac{18}{98} \times 100 \% \sim 18\%$.

\section{Free-rider Game in Federated Learning for Wireless Signal Classification} \label{sec:Wireless}
The game formulation is applied to the case when FL is used to collectively train a \emph{wireless signal classifier}. Each client corresponds to a \emph{spectrum sensor} and collects the in-phase/quadrature (I/Q) data over the air. Each client trains a convolutional neural network (CNN) model to classify the received signals according to their modulation types and coordinates with other clients to collective train a CNN model at the server.  The input to the CNN is of two dimensions (2,16) corresponding to 16 I/Q samples. The output labels are the modulation types, in particular BPSK and QPSK.  Each client has 1000 data samples. Signals are subject to Additive White Gaussian Noise (AWGN) such that the signal-to-noise-ratio (SNR) is uniformly randomly distributed between 0 dB to 10 dB. In addition, phase shifts that are uniformly randomly selected from $[-\pi/30, \pi/30]$ are added to signals. These phase shifts are updated at every 20 samples.  The data samples are split into 80\% and 20\% to construct the training and test datasets, respectively. 

The utility reflects the server accuracy as the reward that measures the accuracy of the global model trained by FL and tested over the separate set of test samples. During FL, each sensor trains its own CNN model (with architecture shown in Table~\ref{table:DNN}) for signal classification. The categorical cross-entropy is used as the loss function and Adam is used as the optimizer. The numerical results are obtained in Python and the CNN models are trained in Keras with TensorFlow backend. FL is applied with FedAvg for model aggregation and run for 100 epochs to measure the global model accuracy.     
\begin{table}
	\caption{CNN architecture used for wireless signal classification.}
	\centering
	{\small
		\begin{tabular}{l|l}
			Layer & Properties \\ \hline \hline
			 Conv2D & filter size = 32, kernel size = (1,3),  \\ & activation = ReLU \\ \hline
			 Flatten & -- 	 \\ \hline
			 Dense & size = 32, activation = ReLU	 \\ \hline
			 Dropout & dropout ratio =  0.1	 \\ \hline
			 Dense & size = 2, activation = Softmax	 \\ \hline
		\end{tabular}
	}
	\label{table:DNN}
\end{table}

\begin{figure}[t]
	\centering
	\vspace{-0.3cm}
	\includegraphics[width=\columnwidth]{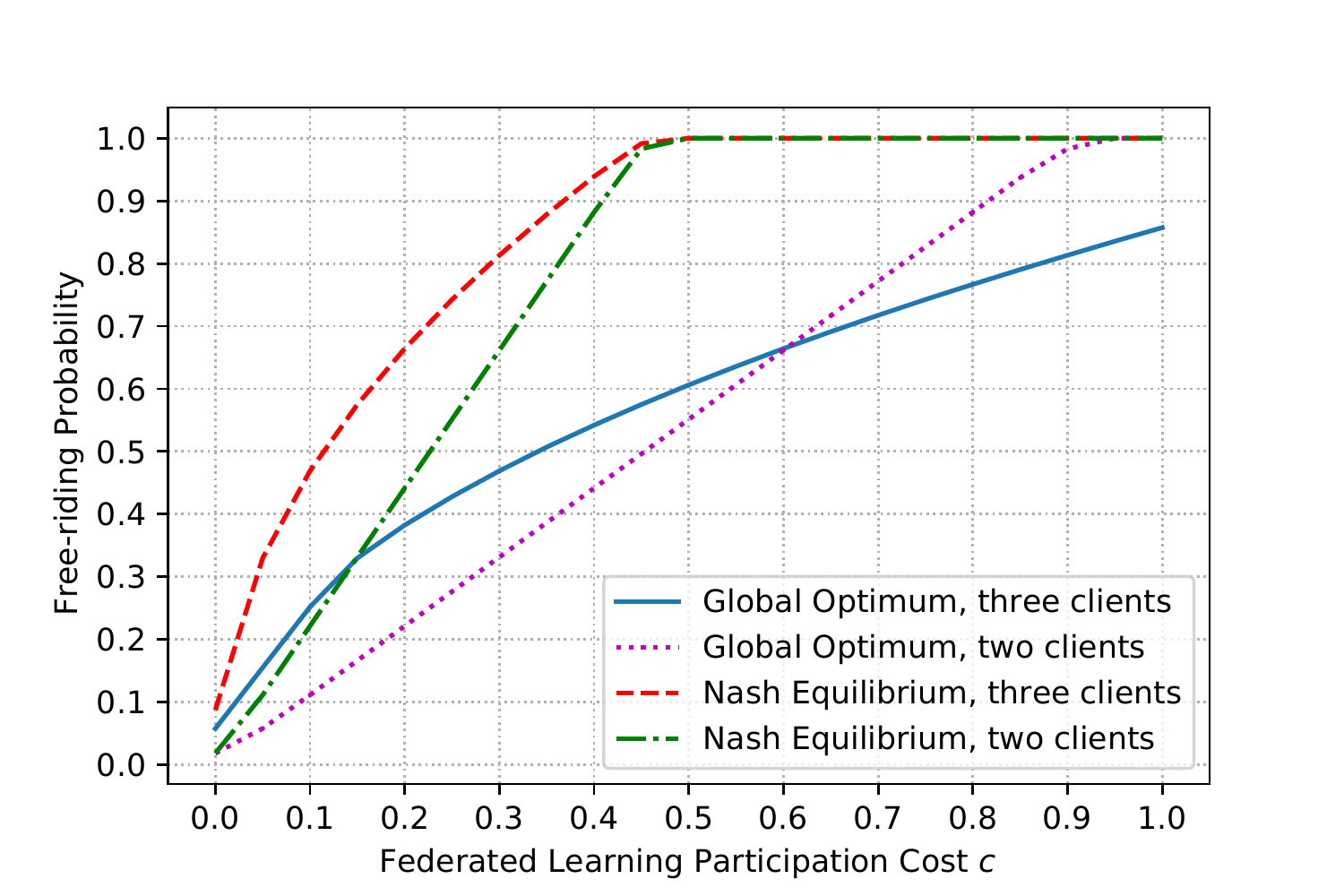}
	\caption{The free-riding probabilities in Nash equilibrium and globally optimal case as a function of $c$ for two and three clients.}
	\label{fig:probFL23}
\end{figure}

\begin{figure}[t]
	\centering
	\vspace{-0.5cm}
	\includegraphics[width=\columnwidth]{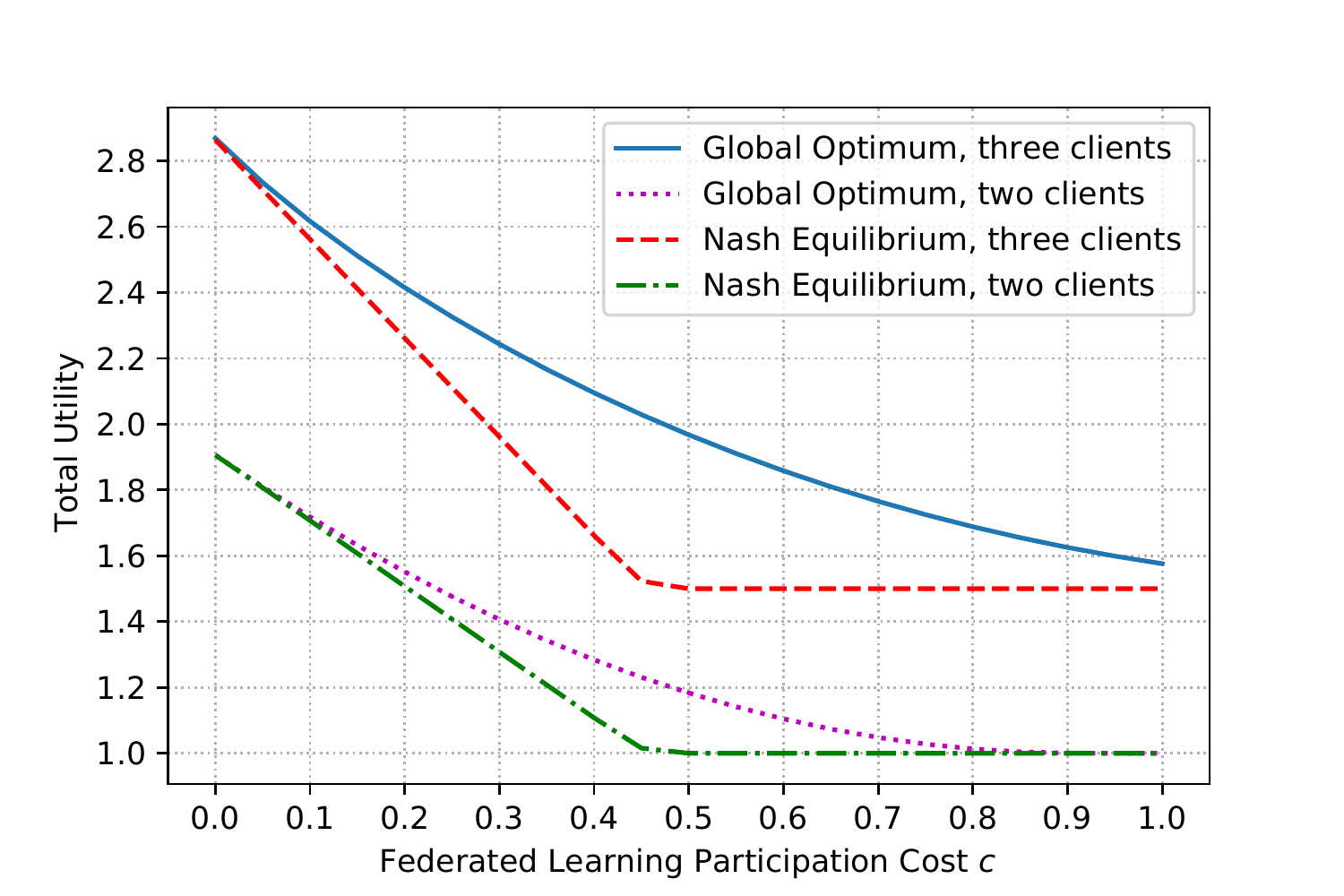}
	\caption{The total utilities in Nash equilibrium and globally optimal case as a function of $c$ for two and three clients.}
	\label{fig:utilityFL23}
\end{figure}

The symmetric strategies are evaluated assuming a common cost $c$. For the case of two and three clients, the symmetric strategies in Nash equilibrium and in the globally optimal case are shown in Fig.~\ref{fig:probFL23} as a function of $c$. The corresponding total utilities are shown in  Fig.~\ref{fig:utilityFL23}. Overall, the highest optimality loss is  $17.87\%$ for two clients and $25.41\%$ for three clients (both observed when $c = 0.46$). Note that as $c$ increases, clients show more tendency to be free-riding (i.e., their free-riding probability increases) such that the total utility drops until it saturates and the free-riding probability becomes $1$. Overall, in Nash equilibrium, the free-riding probability is greater than or equal to the one for the globally optimal case. 

\begin{figure}[t]
	\centering
	\vspace{-0.3cm}
	\includegraphics[width=\columnwidth]{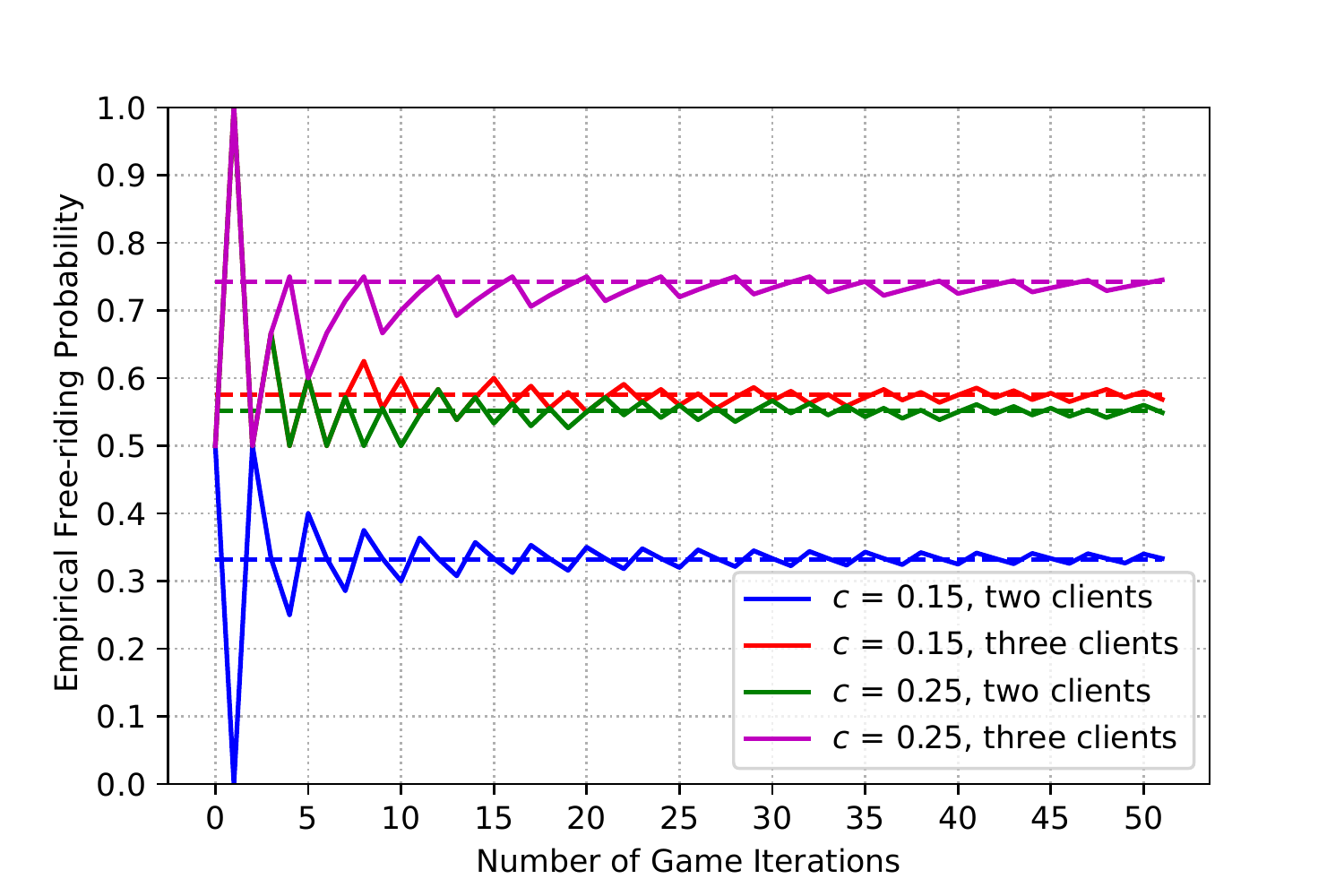}
	\caption{The empirical free-riding probabilities learned over time in fictitious play for two and three clients (dashed straight lines refer to the strategies computed in Nash equilibrium).}
	\label{fig:fictitiousNE23}
\end{figure}

When the \emph{fictitious play} is considered, each client updates its actions by assuming that the other clients play randomized strategies given by the empirical frequency of their past actions. Note that each client does not know the utilities of other clients and only observes an empirical frequency of strategies of other clients. For the first round, each client assumes that the free-riding probability of opponents is 0.5. The empirical free-riding probabilities learned over time for two and three clients are shown in Fig.~\ref{fig:fictitiousNE23}. The fictitious play converges to Nash equilibrium strategies of the corresponding game. 

\begin{figure}[t]
	\centering
	\vspace{-0.5cm}
	\includegraphics[width=\columnwidth]{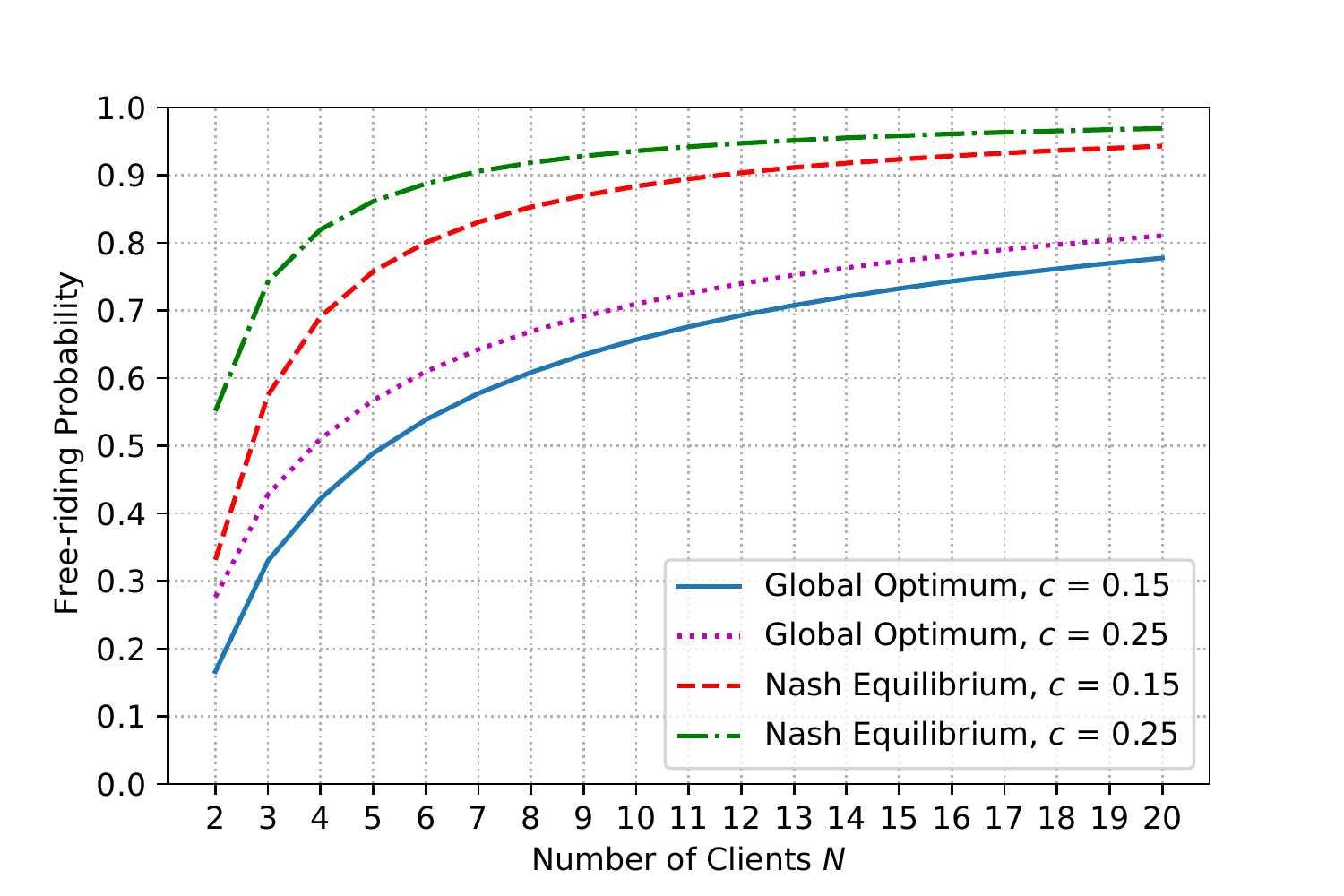}
	\caption{The free-riding probability as a function of the number of clients.}
	\label{fig:RayFLProbNEN20c1525}
\end{figure}

\begin{figure}[t]
	\centering
	\vspace{-0.3cm}
	\includegraphics[width=\columnwidth]{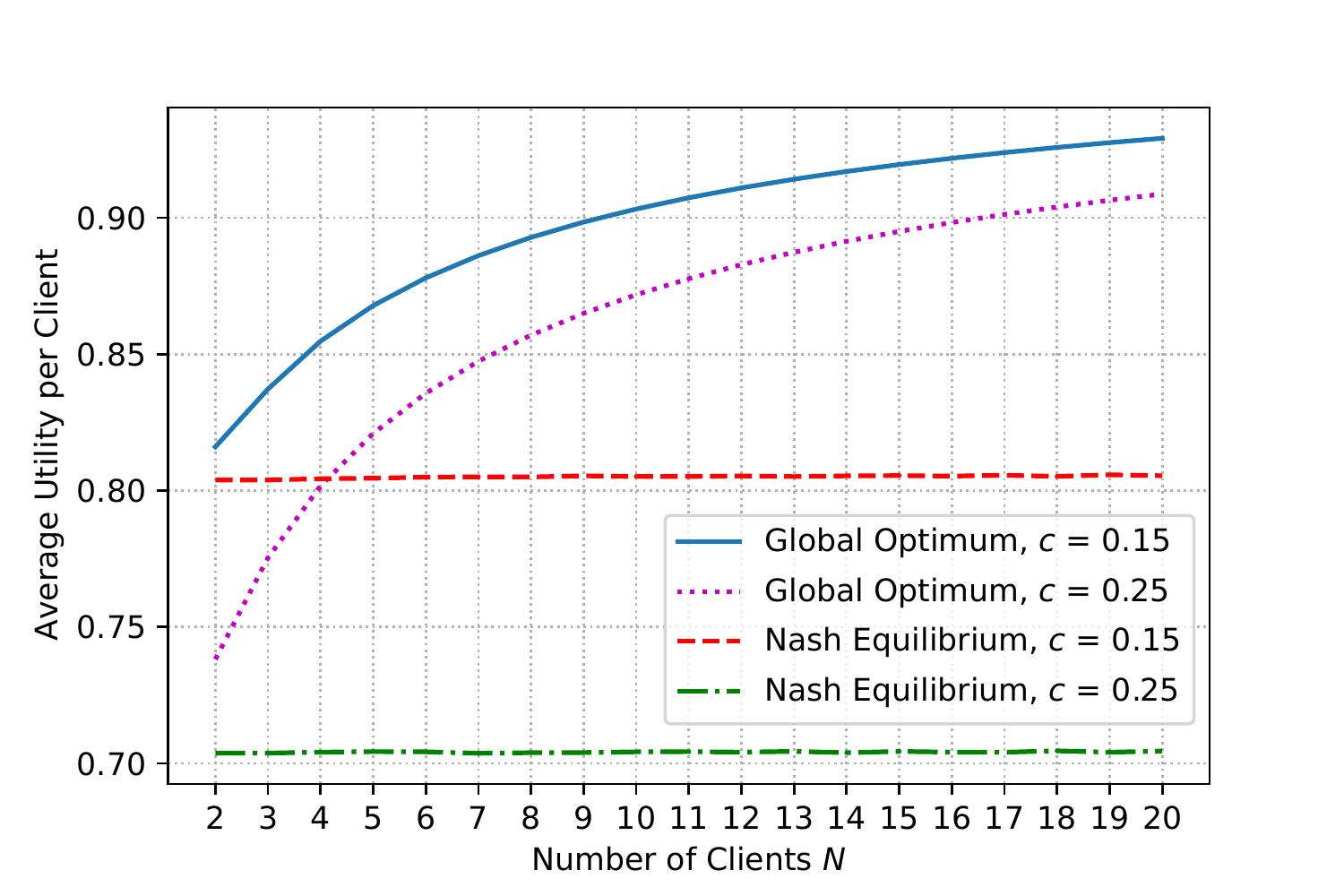}
	\caption{The average utility (per client) as a function of the number of clients.}
	\label{fig:RayFLAvgNEN20c1525}
\end{figure}

Next, the Nash equilibrium strategies and the corresponding utilities are evaluated as the number of clients, $N$, increases. The free-riding probability and the corresponding utility per client as a function of $N$ are shown in Fig.~\ref{fig:RayFLProbNEN20c1525} and Fig.~\ref{fig:RayFLAvgNEN20c1525}, respectively, for $c=0.15$ and $c= 0.25$. As $N$ increases, the free-riding probability increases (sublinearly) and it is higher in Nash equilibrium compared to the globally optimal case. As $N$ increases, the optimality gap increases, since the average utility per client remains constant in Nash equilibrium and increases with $N$ in the globally optimal case. For example, when there are 20 clients involved in FL, the optimality gap becomes $13.32\%$ for $c = 0.15$ and $22.47\%$ for $c = 0.25$. Additional methods such as pricing can be further considered to motivate the clients to act less selfishly and participate more in FL model updates such that the optimally gap can be reduced.

\section{Conclusion}\label{sec:Conclusion}
This paper formulated a game theoretic framework to analyze the free-riding behavior of selfish clients in FL. The free-riding probabilities were derived as the Nash equilibrium strategies when FL is executed over wireless links subject to computation and transmission costs. A non-cooperative game was set up among selfish clients that may assume roles of participating in the FL model updates or free-riding (i.e., receiving the global model without participating). By incorporating the reward (global model accuracy) and the (communication and computation) cost for each client, the utility to be maximized by each client was determined depending on the client's assumed role. The Nash equilibrium strategies were compared to the globally optimal (cooperative) strategies to evaluate the optimality gap. The free-rider game was also set up as the fictitious play, where each client plays the game in response to the empirical frequency of past actions of other clients. This game formulation was applied to the case when FL trains a wireless signal classifier. The results quantified the increase of the free-riding probabilities in Nash equilibrium with the participation cost and the number of clients involved in FL, and characterized the optimality gap that increases with the number of clients. These results indicate operational modes for FL participation by quantifying the impact of free-riding on the resilience of FL in NextG wireless networks.

\bibliographystyle{IEEEtran}

\end{document}